\documentclass[pra, 10pt, twocolumn, letter, floatfix]{revtex4-1}

\usepackage{amsmath}
\usepackage{amsfonts}
\usepackage{amssymb}
\usepackage{graphicx}
\usepackage{dcolumn}
\usepackage{bm}
\usepackage{subcaption}
\usepackage{grffile}

\begin{document}

\title[Real Space Number Counting Jastrows] {Suppressing ionic terms with number counting Jastrow factors in real space}

\author{Brett Van Der Goetz$^{1}$}
\author{Eric Neuscamman$^{1,2,}$}
\email{eneuscamman@berkeley.edu}
\affiliation{$^1$Department of Chemistry, University of California, Berkeley, California 94720, USA\\
             $^2$Chemical Sciences Division, Lawrence Berkeley National Laboratory, Berkeley, California 94720, USA}

\date{\today}

\begin{abstract}
We demonstrate that 4-body real space Jastrow factors are, with the right type of Jastrow basis function,
capable of performing successful wave function stenciling to remove unwanted ionic terms
from an overabundant fermionic reference without unduly modifying the remaining components.
In addition to greatly improving size consistency (restoring it exactly in the case of a geminal power),
real-space wave function stenciling is, unlike its Hilbert space predecessors, immediately compatible with
diffusion Monte Carlo, allowing it to be used in the pursuit of compact, strongly correlated trial functions
with reliable nodal surfaces.
We demonstrate the efficacy of this approach in the context of a double bond dissociation by using it to
extract a qualitatively correct nodal surface despite being paired with a restricted Slater determinant,
that, due to ionic term errors, produces a ground state with a qualitatively incorrect nodal surface when
used in the absence of the Jastrow.
\end{abstract}

\maketitle

\section{Introduction}
\label{sec:introduction}

Linear wave functions in quantum chemistry are fundamentally limited by their inability to compactly express wave functions
in strongly correlated regimes, a difficulty that arises directly from the factorial growth of Hilbert space in the quantum many-body problem.
In practice, therefore, the field of quantum chemistry has long pursued sophisticated nonlinear forms for its approximate wave function ansatzes.
\cite{POPLE:1953:agp,ROOS:1987:casscf,RUEDENBERG:1982:casscf,Bartlett:2007:cc_rev,Chan:2011:dmrg_in_chem_rev}
A key challenge arises in this pursuit due to the difficulty of constructing an ansatz that is simultaneously size-consistent (giving the same energy for
independent systems when treated together or individually) and variational (giving an upper bound to the true energy) while maintaining a cost that scales
polynomially with system size.
Recently, wave function stenciling, wich is a generalization of Gutzwiller's approach \cite{Gutzwiller:1963:gf} in which a nonlinear correlation factor
removes unsuitable terms from an overabundant fermionic expansion,
has been shown to achieve these three properties, \cite{Neuscamman:2012:sc_jagp,Neuscamman:2013:hilbert_jagp,Neuscamman:2016:subtractive}
and so appears to be a promising paradigm for future ansatz design.

While the Jastrow Antisymmetric Geminal Power (JAGP) in Hilbert space is characteristic of this approach
and has proven effective at capturing strong correlation during bond dissociations, \cite{Neuscamman:2013:or_jagp,Neuscamman:2016:ScO}
it is much less effective for capturing the full range of dynamic correlation effects.
One way to understand this difficulty is to consider that its number-operator-based Jastrow factor, which is central to its stenciling strategy,
\cite{Neuscamman:2013:hilbert_jagp} can also be written as a very limited coupled cluster doubles operator. \cite{Neuscamman:2013:or_jagp}
Although sufficient for stenciling, this incomplete reproduction of the doubles operator only partially recovers dynamic correlation.
In short, the Hilbert space Jastrow factor is effective at making large changes to the wave function through stenciling, but much less so
at making the multitude of small changes demanded by dynamic correlation.
In contrast, more traditional Jastrow factors in real space, \cite{FouMitNeeRaj-RMP-01} especially when working in tandem with
diffusion Monte Carlo (DMC), \cite{FouMitNeeRaj-RMP-01} are renowned for their ability to capture dynamic correlation.
Indeed, this pairing has been employed as a reliable substitute for the ``gold standard'' coupled cluster with singles, doubles, and
perturbative-triples (CCSD(T)) \cite{Bartlett:2007:cc_rev} in cases where the latter's higher cost scaling makes it untenable.
\cite{Gillan:2012:water_clusters}

This dichotomy between Jastrow factors' strengths in real space and Hilbert space raises the natural question: what is preventing
the development of a Jastrow factor that can deliver both the large changes required for stenciling and the small changes for dynamic correlation?
Further: can these obstacles be overcome in a way that maintains both low-order polynomial scaling as well as suitability for use as
as a DMC guiding function in strongly correlated regimes, where the best current option is to rely on a factorial-cost determinantal expansion?
In this paper, we will present a real space formulation that answers these questions in the affirmative,
explain why previous real space Jastrows were not able to live up to this ideal,
demonstrate that Jastrow-based stenciling can be effective even when the stenciled wave function is a single Slater determinant,
and offer some thoughts on the requirements that should be satisfied in future by a general-purpose stenciling Jastrow.

The essential challenge to performing stenciling in real space is that any attempt to delete large portions of the wave function
using a multiplicative Jastrow factor will require its functional form to contain a high degree of curvature.
Such curvature is necessary, as any smooth function that asymptotes to a constant value at infinite distance (as a Jastrow factor should)
and contains little curvature will be similar to a constant, and multiplication by this nearly-constant function will not produce
large changes in the wave function.

Unless the large curvature needed for stenciling can be hidden in some way, its tendency to raise the kinetic energy
will lead the variational principle to eschew Jastrow-based deletion of undesirable configurations, even in cases where the
functional form could accommodate it.
To address this challenge, we present a new form of four-body Jastrow factor that is better-suited to hiding its curvature in regions of low
wave function value (where it will not affect kinetic energy) and to counting electrons within local regions of space
(the mechanism by which Hilbert space Jastrows achieve stenciling).
Combined with traditional two-body Jastrows, a Slater determinant, and diffusion Monte Carlo, these real space number counting Jastrow
factors allow for an effective description of both static and dynamic correlation within a structure whose complexity is explicitly polynomial.

\section{Theory}
\label{sec:theory}

\subsection{Mimicking Hilbert Space Jastrows}
\label{sec::mimic_hsjf}

Let us begin by reviewing Hilbert Space Jastrow Factors (HSJFs), which may be written in terms of a matrix $\bm{F}$
and the second quantized number operators $\hat{n}_i$ within an orthonormal (and typically local) one particle basis,
\begin{align}
\label{eqn:hsjf_general_form}
e^{\hat{J}_{HS}} = \exp\left( \sum\limits_{ij} F_{ij} \hat{n}_i \hat{n}_j \right) = \prod\limits_{ij} \exp\left( F_{ij} \hat{n}_i \hat{n}_j \right).
\end{align}
Note that these can be thought of as four-body e-e-N-N Jastrow factors, as the indices $i$ and $j$ run over orbitals that
are localized at or near the nuclei while the results of operating with the number operators tell us about the positions
of up to two different electrons.

As number operators are idempotent and overall constant factors irrelevant, $\bm{F}$ can be chosen such that
the HSJF contains any number of Gaussian factors
\begin{align}
\label{eqn:gauss_factor}
\hat{\Gamma}_{W,\mathrm{\hspace{0.6mm}HS}} = \exp \left( -\xi\left( N - \sum\limits_{p\in W}\hat{n}_p \right)^2 \right)
\end{align}
for use in wave function stenciling.
Application of one of these factors to a fermionic wave function effectively reweights each configuration in that wave function's expansion
within this particular orbital basis according to a Gaussian distribution in the total occupancy of an orbital subset $W$.
Provided that the ``projection strength'' $\xi$ is sufficiently large, such a Gaussian factor acts as a stencil, removing any configuration
in which the set of orbitals $W$ contains an electron count differing from $N$.
Given two or more molecular fragments, this effect can be used to eliminate any configurations in which a fragment possesses an
unphysical charge (an "ionic configuration"), which turns out to be sufficient for restoring size consistency to a geminal power \cite{Neuscamman:2012:sc_jagp}.
Crucially, this factor does nothing to components of the wave function which do not deviate from the prescribed pattern of subsystem electron counts,
thus preventing the HSJF from raising the kinetic energy of configurations that survive the stencil.

Unfortunately, a direct translation of the HSJF into real space is problematic for QMC methods due to the nonlocal nature
of a number operator's real space form,
\begin{align}
\label{eqn:real_space_n}
\hat{n}_p = \sum\limits_i \phi_p(r_i) \int \phi_p^*(\tilde{r}_i) d\tilde{r}_i.
\end{align}
Efficient stochastic interrogations of a wave function in real space hinge on the ability to evaluate local wave function values
$\Psi(\vec{r})$, which is complicated by the number operators' nonlocality.
Instead, we will seek a local function $Q$, associated with a region $W$, enclosed within a Jastrow factor of similar Gaussian form
\begin{align}
\label{eqn:gauss_with_c}
\Gamma_{W,\mathrm{\hspace{0.6mm}RS}} = \exp\left( -\xi\left( N - \sum\limits_{i} Q(r_i)\right)^2\right)
\end{align}
that permits efficient local evaluation and, thanks to the sum over all the electron positions $r_i$, maintains the bosonic symmetry
required by the Jastrow factor to keep the overall wave function correctly antisymmetric.

In order to mimic the effects of a HSJF, we therefore desire that each real space Gaussian component approximate the effects of
its Hilbert space counterpart as closely as possible at any sampled position of the electrons; thus we want
\begin{align}
\label{eqn:match_gaussians}
\langle r | \Gamma_{W,\mathrm{\hspace{0.6mm}RS}} | \Psi \rangle \simeq \langle r | \hat{\Gamma}_{W,\mathrm{\hspace{0.6mm}HS}} | \Psi \rangle,
\end{align}
where $|\Psi\rangle$ is the fermionic wave function that is to undergo stenciling.
When basis orbitals in $W$ are spatially separated from others in the system
\textemdash\ an ideal that is often approached in the localized physics of strong correlation \textemdash\ it is sufficient
to choose $Q$ as a step function:
\begin{align}
\label{eqn:step_func}
Q(r_i) \rightarrow \left\{ \hspace{2mm} \begin{matrix} 1 & r_i \in R_W \\ 0 & r_i \not\in R_W \end{matrix} \right.
\end{align}
in which $R_W$ is a region exclusively supporting the orbitals in $W$.
To preserve smooth wave function derivatives and allow for a gradual approach to step like behavior in cases where orbital subsets
are partially overlapping in space, we relax the step discontinuity at the boundaries of $R_W$ by employing an analytical
approximation to the Heaviside function (see Section \ref{sec::ncjf_form}).

So long as the smoothed form of Q rapidly approaches 0 as one moves away from the boundary of $R_W$, the Jastrow factor of
Eq.\ (\ref{eqn:gauss_with_c}) retains the ability to precisely control the electron count on a subsystem that is spatially
well-separated from other subsystems, as there is in this case ample room in between for the 1-to-0 switch to occur.
Thus, as with a HSJF, the real space form presented here can fully eliminate ionic terms between well-separated subsystems,
allowing it to restore exact size consistency to geminal powers and to aid in the repair of restricted Slater determinants.
The key question now becomes whether we can construct functional forms for $Q$ that permit useful demarcations of spatial regions
while also ensuring that the curvature they introduce can be hidden in regions where its contribution to the kinetic energy, through the term
\begin{align}
\label{eqn:step_func}
-\frac{1}{2} \int dr \langle\Psi|r\rangle \langle r | \big( \nabla^2 \Gamma_{W,\mathrm{\hspace{0.6mm}RS}} + 2\nabla \Gamma_{W,\mathrm{\hspace{0.6mm}RS}} \cdot \nabla \big) |\Psi \rangle,
\end{align}
is mitigated by small local wave function amplitudes $\langle\Psi|r\rangle$.
This is of course trivial when demarcating a region around a well-separated fragment, but becomes less so
during dissociation events, where partial stenciling becomes beneficial long before the well-separated limit is reached.

\subsection{Existing 4-body Jastrows}
\label{sec::existing_jastrows}

Before detailing our proposed form for a stenciling-friendly 4-body Jastrow factor, it is instructive
to consider why existing 4-body forms are ill-suited for this task.
Begin by considering a previously used form \cite{Sorella:2009:rvb_molecules} for 4-body Jastrows that closely mirrors that of a HSJF:
\begin{align}
\label{eqn:4body_form}
e^{J_4} = \exp\left( \sum\limits_{ijIJ} \Phi_I(r_i) F_{IJ} \Phi_J(r_j)  + \sum\limits_{iI} G_I\Phi_I(r_i) \right).
\end{align}
By diagonalizing $\bm{F}$, choosing $\vec{G}$ appropriately, and ignoring changes to wave function normalization,
one may convert this Jastrow into a product of Gaussians,
\begin{align}
\label{eqn:4body_guass_form}
e^{J_4} \rightarrow \prod_J \exp\left( -\xi_J \left( N_J - \sum\limits_{iI} U_{IJ} \Phi_I(r_i)\right)^2\right),
\end{align}
in which $\bm{U}$ is the unitary matrix that diagonalizes $\bm{F}$.
Written this way, we may immediately identify the linear combination $\sum_I U_{IJ} \Phi_I$
as one possible form for the counting function $Q$ discussed in the previous section.

We may evaluate the suitability of 4-body Jastrows of the type given in Eq.\ (\ref{eqn:4body_form}) for use in HSJF-style stenciling
by asking how easily these linear combinations can approximate a step function over a given region,
and how much control they have over their curvature.
By considering the task of controlling the electron count on a single atom well-separated from the remainder of whatever system is
being modeled, the above analysis makes plain that the two common forms for the basis functions $\Phi_I$,
atom-centered Gaussians \cite{gaussian_jastrow} and symmetric polynomials \cite{polynomial_jastrow},
are not effective for wave function stenciling in Hilbert Space.
In the same way that one requires many Fourier components to converge to a square wave, small Gaussian expansions or low-order polynomial
expansions are unable to faithfully approximate the switching behavior required for our Jastrow basis functions.
Indeed, Gaussian functional forms contain significant curvature at and about the atom's center where the wave function is large in magnitude,
and thus cannot engage in the curvature hiding necessary to avoid a rise in kinetic energy when $\xi_J$ is large, i.e.\ in the
strong stenciling regime.
Although it is true that in the infinite basis set limit, a complete set of functions (such as the Gaussian spherical harmonics) can represent
any smooth function, they will converge to the nearly steplike behavior required by $Q$ only very slowly and so will
retain appreciable curvature near the center of the counting region unless the Jastrow basis is made extremely large.
In practice, therefore, the functional forms within previously-studied 4-body Jastrows were inappropriate
for stenciling, and so, during optimization, the variational principle did not explore their ability to eliminate ionic
terms, as doing so would have led to large, curvature-induced increases in the kinetic energy.
Ultimately, as can be seen in Sorella's carbon dimer results \cite{Sorella:2007:weak_binding}, the price for using a Jastrow basis that cannot easily
represent a step function is, in the context of the JAGP, a size consistency error stemming from the inadequate suppression of ionic terms.

\subsection{Mathematical formulation}
\label{sec::ncjf_form}

We investigate the efficacy of real space number-counting Jastrow factors (NCJFs) that can be written in the same general structure as existing four-body Jastrows,
\begin{align}
\label{eqn:jcjf_4body_form}
e^{J_{NC}} = \exp\left( \sum\limits_{ijIJ} C_I(r_i) F_{IJ} C_J(r_j) + \sum\limits_{iI} G_I C_I(r_i) \right).
\end{align}
As discussed above, the key characteristic of NCJFs will lie in the choice of basis functions $C_I$, for which we select a form that
can, to a certain degree, act as local real space approximations to Hilbert Space number operators. 
In the limit of disjoint orbital subspaces, bosonic step functions in real space can exactly reproduce the effects of a sum of
Hilbert space number operators and can thus serve as a conceptual starting point for our basis functions.
Although we will soften this step-function extreme by employing smooth functions, we will retain the spirit of
spatially localized curvature so as to facilitate the curvature hiding that NCJFs require in order to effect strong stenciling
without unphysically affecting the kinetic energy.

This goal in mind, we propose ``counting'' basis functions of the form
\begin{align}
\label{eqn:count_func}
C(\mathbf{r}) = S(g(\mathbf{r})),
\end{align}
where the Fermi-Dirac-like function
\begin{align}
\label{eqn:count_func}
S(x) = \frac{1}{1+e^{-\beta x}}
\end{align}
plays the role of an analytic approximation to the Heaviside step function.
The value of $S$ asymptotically switches from zero to one as its argument traverses the origin,
with $\beta$ (which is \textit{not} related to physical temperature) determining both the slope at the origin as well as
the effective width of the switching region in which $S$ meaningfully differs from zero or one and displays non-negligible curvature.
The interior function $g(r)$ is a scalar-valued function of a real-space coordinate whose nodal surface defines the boundary,
or switching surface, of the region within which electrons are to be counted.
The volume for which $g$ takes on positive values (negative values) is called the interior (exterior) of the counting region,
since composition with the switching function $S$ ensures that $C$ asymptotically evaluates to one (zero) inside this region. 
We will refer to counting functions by the geometry of their switching surface, and in the present study we investigate
both planar
\begin{align}
\label{eqn:planar_g}
g_P(\mathbf{r}) = (\mathbf{r} - \mathbf{c}) \cdot \hat{k}
\end{align}
and elliptical 
\begin{align}
\label{eqn:elliptical_g}
g_E(\mathbf{r}) = (\mathbf{r} - \mathbf{c})^T T(\mathbf{r}-\mathbf{c}) - 1
\end{align}
counting regions.
The nodal surface of $g_P$ is a plane centered at $\mathbf{c}$ and normal to the unit vector $\hat{k}$,
while the nodal surface of $g_E$ is an ellipsoid with center $\mathbf{c}$ and axes defined by the
eigenvectors and eigenvalues of $\bm{T}$.
Together with $S$, these counting regions provide us with a set of Jastrow basis functions whose only curvature
appears at the edges of their counting regions, making it more amenable to being hidden in regions of low wave function
magnitude.
This positioning of curvature should be compared to the more traditional forms given in Section \ref{sec::existing_jastrows},
which display significant curvature at their centers.

In addition to this stenciling-friendly curvature, arithmetic operations between these counting basis functions correspond to
set operations between their interior volumes, which gives their sums and products a somewhat intuitive meaning.
For example, consider the large-$\beta$ limit of these counting functions,
\begin{align}
\label{eqn:large_beta_limit}
\lim_{\beta\rightarrow\infty} C(\mathbf{r}) = \left\{ \hspace{2mm} \begin{matrix} 1 & \mathbf{r} \in A \\ 0 & \mathbf{r} \not\in A \end{matrix} \right.
\end{align}
in which these functions revert to actual step functions.
Spatial regions' complements now occur simply as
\begin{align}
\label{eqn:compliment}
1 - C_A(\mathbf{r}) = C_{A^c}(\mathbf{r}),
\end{align}
while intersections
\begin{align}
\label{eqn:intersections}
C_A(\mathbf{r}) C_B(\mathbf{r}) = C_{A \cap B}(\mathbf{r})
\end{align}
and unions
\begin{align}
\label{eqn:unions}
C_A(\mathbf{r}) + C_B(\mathbf{r}) - C_{A\cap B}(\mathbf{r}) = C_{A\cup B}(\mathbf{r})
\end{align}
arise from products and sums of counting functions.
In this way, the quadratic form in Eq.\ (\ref{eqn:jcjf_4body_form})
offers the possibility for the full set of first-order topological operations to arise naturally during the
variational minimization of a NCJF, raising interesting questions as to whether adjacent regions will
merge or produce cutouts from one another in pursuit of optimal stencils.

\section{Results}
\label{sec:results}
We have prepared a pilot implementation supporting planar and elliptical NCJFs within
a development version of QMCPACK. \cite{Kim:2012:qmcpack_scaling}
NCJFs, as well as spline-based, cusp-correcting $e$-$e$ and $e$-$n$ two-body Jastrows and the molecular orbital coefficients
were optimized with respect to energy using the variational Monte Carlo (VMC) linear method
\cite{Nightingale:2001:linear_method,UmrTouFilSorHen-PRL-07,TouUmr-JCP-07,TouUmr-JCP-08}.
The Hamiltonian is taken as the non-relativistic electronic Hamiltonian under the Born-Oppenheimer approximation,
with effective core potentials \cite{carbon_pseudopotential} used to replace carbon atoms' core electrons.
RHF solutions are taken as the reference configurations at each geometry in the cc-pVTZ basis \cite{cc-pvdz} and are
generated by GAMESS\cite{gamess}.
Multireference configuration interaction calculations with the Davidson correction (MRCISD+Q) were performed with
\uppercase{MOLPRO} \cite{MOLPRO_brief} also in the cc-pVTZ basis.

\subsection{Nomenclature}

In our results, we will distinguish wave functions based on the types of Jastrow factors employed, whether or not
the molecular orbitals were re-optimized in the presence of the Jastrow, and, where applicable, whether the 
molecular orbitals are symmetric (SA) or have broken symmetry (SB).
The presence of counting Jastrows will be denoted by C, traditional spline-based $e$-$e$ and $e$-$n$ Jastrows by T,
and orbital re-optimization by the prefix ``oo-''.
In all cases, JS stands for Jastrow-Slater.
For example, a Jastrow-Slater wave function with both traditional and counting Jastrows whose orbitals were
re-optimized starting from a broken symmetry orbital guess would be denoted as oo-CTJS-SB.
Finally, DMC results will be denoted by adding DMC to the name of the wave function that fixes the nodal surface.

\subsection{Hydrogen Molecule}

As a minimally correlated wave function, a single restricted Slater determinant is insufficient to describe electron correlation at stretched molecular geometries,
which can lead to large size-consistency errors during molecular fragmentation.
To correct this, we apply a simple NCJF with a basis consisting of two anti-aligned planar counting functions whose switching surfaces are set to bisect the H-H bond. 
The NCJF matrix parameters $F_{IJ}$ and $G_I$ are initially set to zero (so that the overall Jastrow factor is initially unity), after which both the matrix parameters
($F_{IJ}$, $G_I$) and basis function parameters ($\beta_I, \hat{k}_I, \mathbf{c}_I$) are optimized.
Figures 1 and 2 show that NCJFs paired with either cusp-correcting Jastrows (CTJS), orbital optimization (oo-CJS), or both (oo-CTJS) prove far more effective
at recovering size-consistency than when only two-body Jastrows (TJS) are used, even if assisted by orbital re-optimization (oo-TJS).
At a separation of 4 \AA, for example, we find that oo-CTJS is size-consistent to within 0.4 mE$_h$,
while the smallest size-consistency error achievable without NCJFs is over 14 mE$_h$.

\subsection{Ethene}

A variationally optimized Slater-Jastrow ansatz is often taken as a guiding function for diffusion Monte Carlo calculations, but the appearance of symmetry-broken
minimum-energy solutions to the RHF equations at stretched geometries --- which do not possess the correct nodal structure required by DMC --- means that we cannot
naively take minimum-energy configurations without issue.
For instance, when stretching the C=C bond in ethene past 2.5 \AA, an RHF solution with broken-symmetry orbitals sees its RHF energy drop below that of the
symmetric-orbital solution.
However, the nodal surface of this broken-symmetry solution is incorrect, and so when used in DMC it gives an energy that is 40 mE$_h$ or more above
that of a DMC based on the symmetric-orbital RHF solution (see Figure \ref{fig:figure3}).
In more complicated systems, such effects can be more pronounced, and it would be highly desirable to be able to predict beforehand which nodal surface
is most appropriate.
Given a sufficiently flexible trial function to optimize, VMC can in principle produce the correct nodal surface by selecting the VMC wave function with
the lowest energy.
\begin{figure}
	\includegraphics[width=0.99\linewidth]{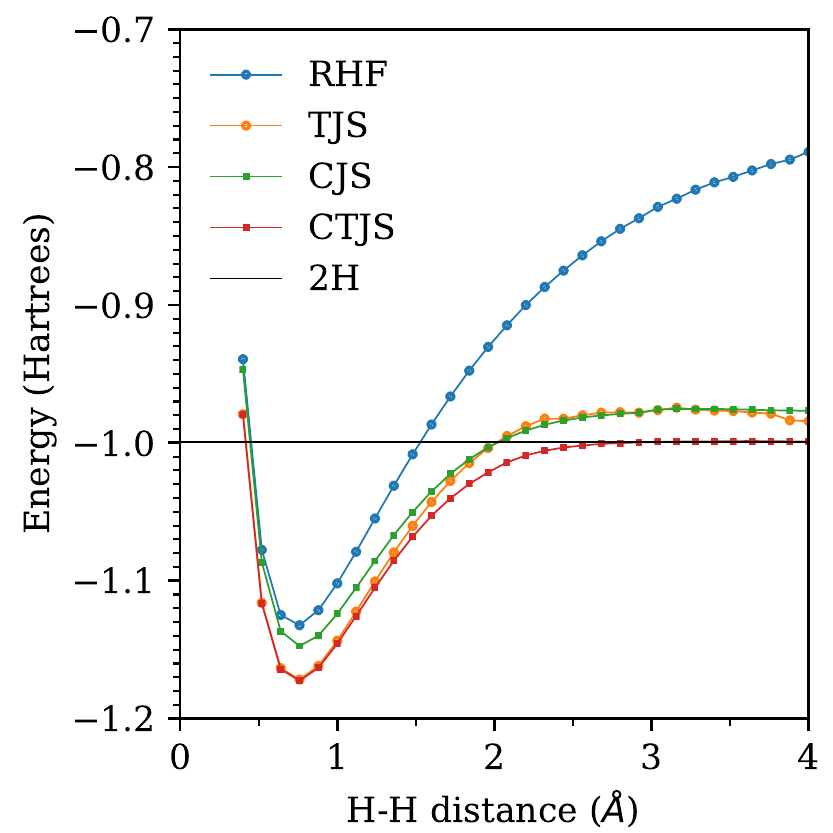}
	\caption{VMC energies for the dissociation of H$_2$ using a cc-pVTZ orbital basis. The solid black line is twice the VMC energy of a single H atom in the same cc-pVTZ basis.}
	\label{fig:figure1}
\end{figure}
\begin{figure}
	\includegraphics[width=0.99\linewidth]{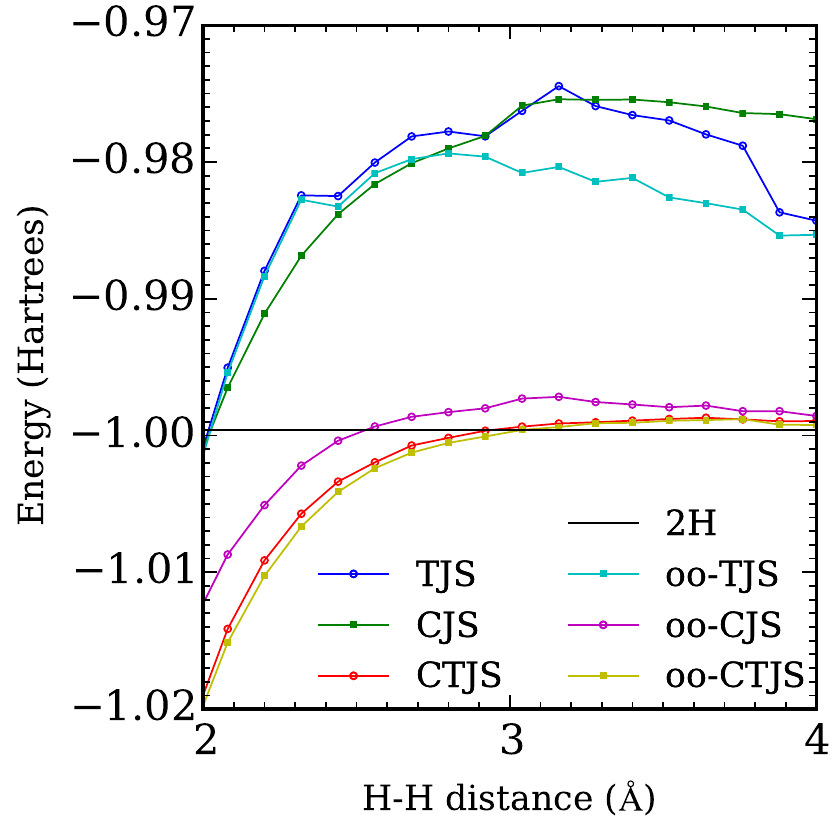}
	\caption{VMC energies for the dissociation of H$_2$ using a cc-pVTZ orbital basis,
           now focusing on stretched geometries.
           The solid black line is twice the VMC energy of a single H atom in the same cc-pVTZ basis.}
	\label{fig:figure2}
\end{figure}
However, this approach will only be reliable if the trial function is flexible enough, and in the case of ethene, traditional Jastrow-Slater is not,
even under orbital re-optimization, as can be seen in Figure \ref{fig:figure4}.
Although multiconfigurational expansions can be used in lieu of a single reference fermionic function in order to achieve the flexibility needed
to describe the strong correlation responsible for flipping the energy ordering of these two states, the complexity and thus cost of such an expansion
must grow exponentially with the number of correlated bonds.
The cost of a stenciling approach using NCJFs --- assuming a constant number of counting basis functions per fragment --- will by comparison scale only
quadratically with fragment number, and so it would be quite useful if stenciling were able to capture a sufficient amount of strong correlation to produce
the correct energy ordering of states at the VMC level.

Using the same planar NCJFs as in the hydrogen case (except now the planes bisect the C=C bond), we apply NCJFs, orbital optimization, and
traditional two-body Jastrows to a single Slater determinant that is either a symmetry-adapted (SA) or symmetry-broken (SB) RHF solution.
(Note that SA vs SB orbitals did not interconvert under orbital optimization and appear to represent two separate minima on the optimization surface.)
In the most flexible case, oo-CTJS, VMC is now correctly able to predict that the SA energy lies below that of the SB energy,
a prediction that fails to materialize if the NCJF is omitted (see Figure \ref{fig:figure4}).
Upon using the oo-CTJS-SA and oo-CTJS-SB wave functions to fix the DMC nodal surface, we find that the lower-energy VMC
state now corresponds to the lower energy DMC result, and that the lower energy DMC result is in close agreement with MRCISD+Q
(see Figure \ref{fig:figure5}).
Thus, in the case of the ethene double bond dissociation at least, the NCJFs' ability to suppress spurious ionic terms
within a single Slater determinant is sufficient to produce a qualitatively correct state ordering and nodal surface without
resorting to multi-determinantal expansions.

\begin{figure}[t]
	\includegraphics[width=1.00\linewidth]{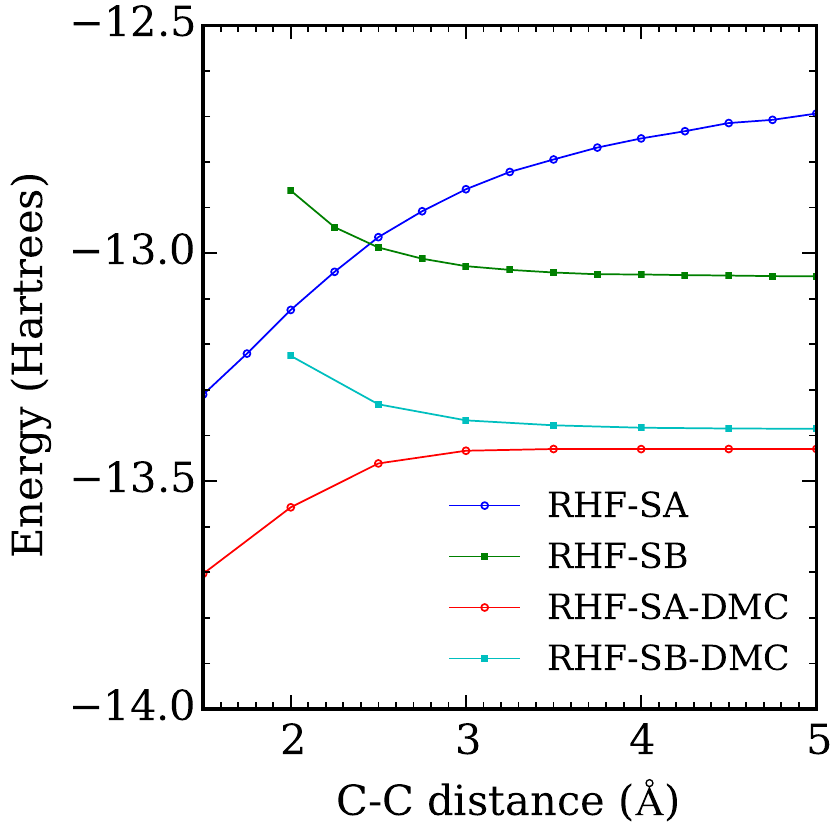}
	\caption{VMC and DMC energies of symmetry-adapted (SA) and symmetry-broken (SB) RHF solutions.
           The RHF minimum at stretched geometries corresponds to the symmetry-broken configuration,
           which is not the best DMC guiding function.}
	\label{fig:figure3}
\end{figure}

\begin{figure}
	\includegraphics[width=1.00\linewidth]{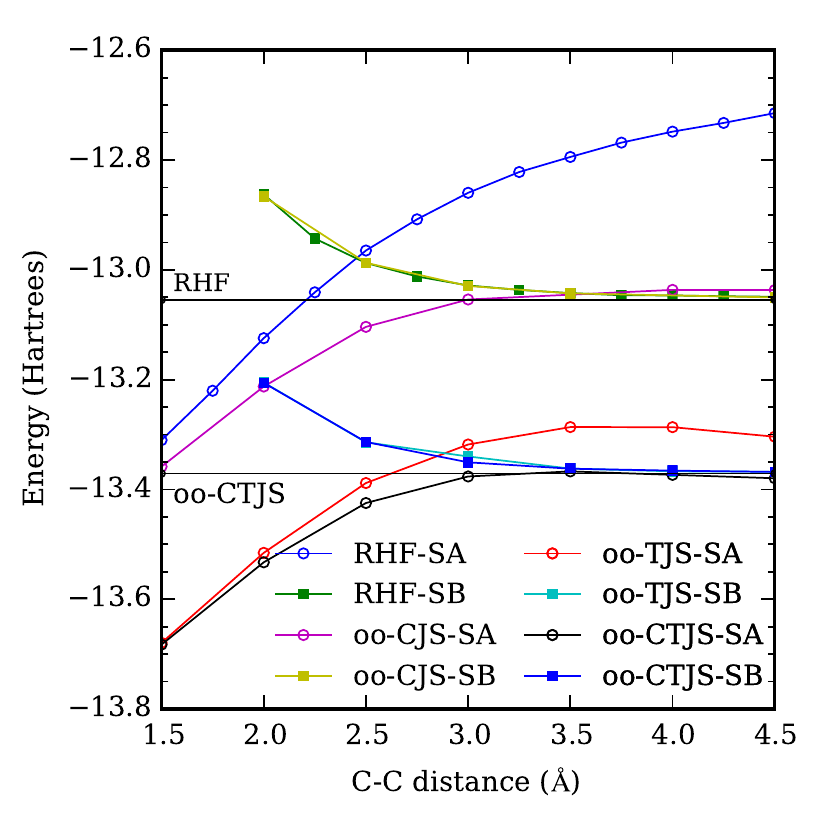}
	\caption{VMC energies with orbitals optimized together with Jastrow variables.
           The solid black lines are twice the VMC energy of a single CH$_2$ fragment,
           with no Jastrow factor (RHF) or with a cusp-correcting Jastrow (TJS),
           as indicated, and provide a reference for size-consistency.}
	\label{fig:figure4}
\end{figure}

\begin{figure}
	\includegraphics[width=0.99\linewidth]{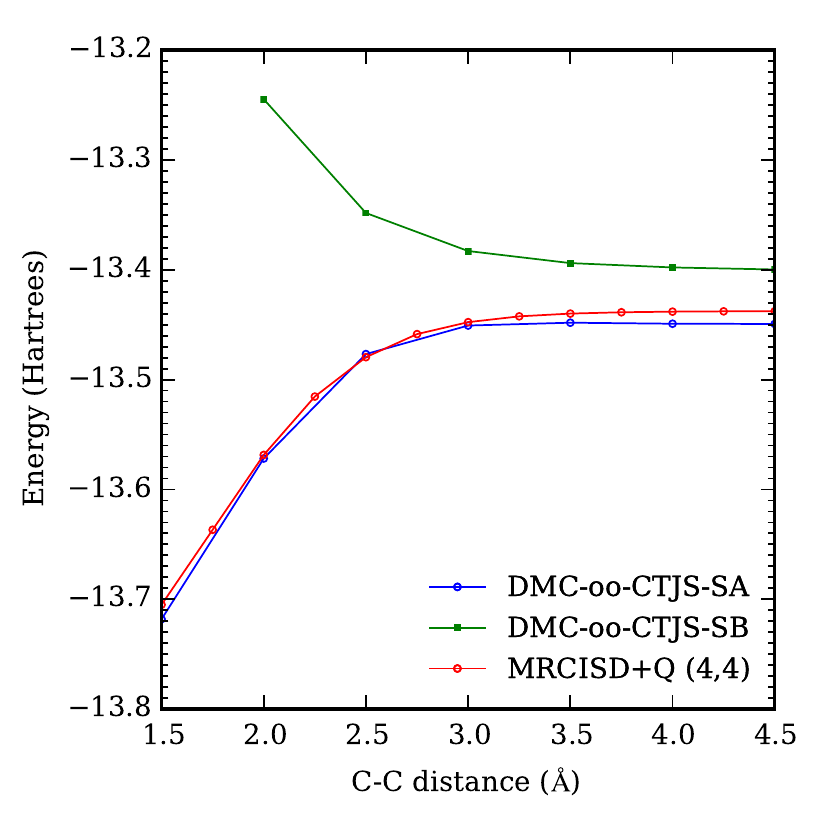}
	\caption{DMC energies based on the oo-CTJS-SA and oo-CTJS-SB nodal surfaces,
           along with results for (4e,4o)-MRCISD+Q.}
	\label{fig:figure5}
\end{figure}

\subsection{Curvature Hiding}

\begin{figure}
	\includegraphics[width=0.99\linewidth]{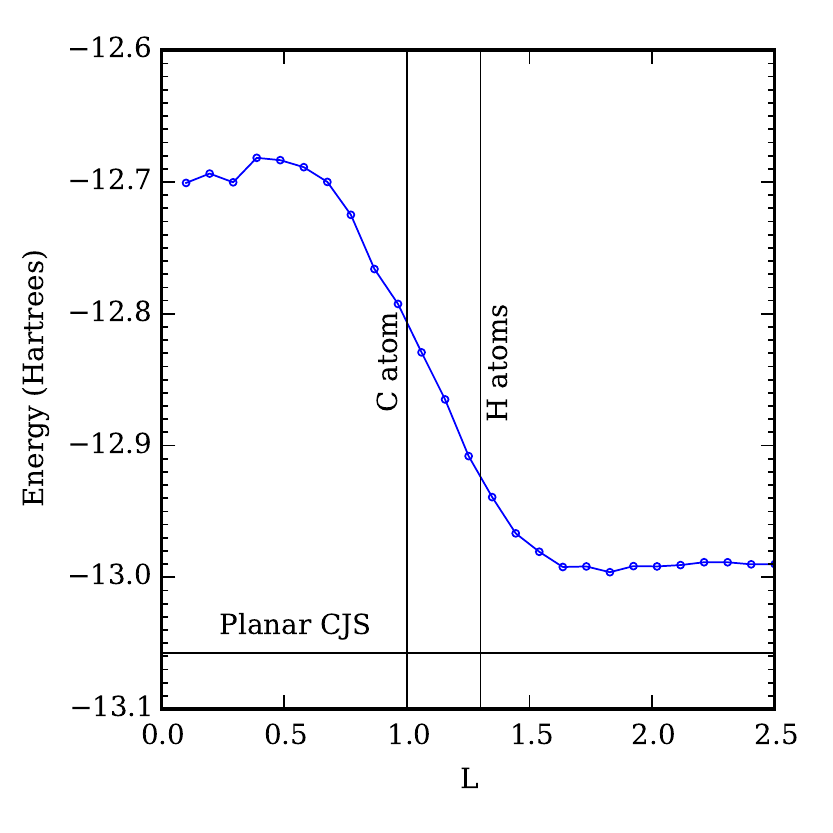}
	\caption{CJS energies using elliptical counting functions of variable size.
           The ellipses always have one edge touching the C=C bond midpoint and lie entirely between the carbons at $L=0$.
           As $L$ is increased, the ellipses grow in size until eventually encompassing an entire CH$_2$ fragment, with vertical lines
           showing the values of $L$ at which the ellipses' outer surfaces cut directly across an atomic nucleus.
           The horizontal black line indicates the VMC energy of planar counting regions, which should serve as a lower bound for the energy in the $L\rightarrow \infty$ limit.}
	\label{fig:figure6}
\end{figure}

Using elliptical basis functions, we will demonstrate the size-consistency problem encountered with basis functions used previously in four-body Jastrow factors.
As noted earlier, the main problem associated with taking atom-centered Gaussian functions as the Jastrow basis lies in their inability to effectively hide their
curvature in regions where wave function values are small.
We reproduce this effect in ethene at a stretched geometry (4.5 \AA) by changing our Jastrow basis from C=C bond-bisecting planes to a set of elliptical counting
functions in which we scan over an axis scaling parameter $L$ while keeping one elliptical edge orthogonally bisecting the C=C bond axis.
NCJF parameters ($\bm{F}$, $\vec{G}$, $\beta$) are fixed at values optimal for planar counting functions, which allows us to reproduce the behavior of
the planar basis as we increase the axis scale (at large values of $L$).
This gives us a suitable baseline to compare the elliptical and planar Jastrow basis at different geometries, and, as we expect, VMC energies match reasonably well
when each elliptical counting function encompasses an entire CH$_2$ fragment.
However, as we shrink the elliptical switching surface to only partially encompass each fragment, Figure \ref{fig:figure6} shows that the overall energy increases,
which can be explained by the fact that the curvature at the edge of the elliptical counting region is now cutting through a region with appreciable wave function
magnitude.
For Jastrow basis functions like these too-small ellipses, the risk of such energetic penalties prevents the variational principle from
allowing the elements of the Jastrow matrix $\bm{F}$ to become large in magnitude, thus precluding any stenciling-like effects.
When instead the basis function allows for curvature hiding, as when the ellipse is large enough so that its edges are outside the boundaries of
the CH$_2$ fragment, the variational principle is free to restore size-consistency by using large-magnitude $\bm{F}$ elements to delete spurious ionic terms.

\subsection{Basis Construction Schemes}

While we have shown that the combination of suitably chosen Jastrow basis functions with a general 4-body Jastrow factor form can successfully
introduce strong correlation effects through the stenciling of ionic terms, a generally applicable method requires some rule or prescription
for how the Jastrow counting functions are to be chosen for an arbitrary molecule.
Let us discuss and discard two options based on the current planes and ellipses before motivating future work with some observations on the
properties that a general NCJF basis should satisfy.

First, one might choose to place planar counting functions so as to bisect each bond in a molecule.
Such an approach would prepare the ansatz for suppressing unwanted ionic terms in any given bond, but the infinite extent of the planes would
clearly not in general satisfy the requirement that the counting function's curvature be hidden in regions of small wave function magnitude.
What if the plane from one bond intersects a far-away atom?

\begin{figure}[t]
	\includegraphics[width=0.99\linewidth]{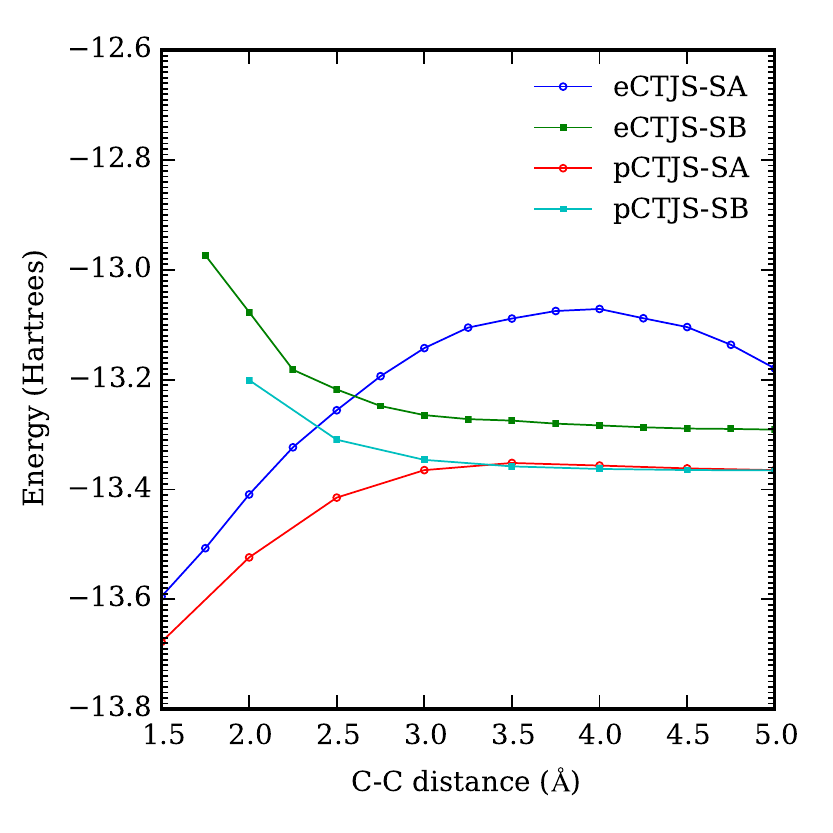}
	\caption{VMC energies when using atom-centered ellipsoids (eCTJS) vs C=C bond bisecting planes (pCTJS) for the counting regions,
           applied to both symmetric and symmetry-broken RHF determinants.}
	\label{fig:figure7}
\end{figure}

Second, one could consider using atom-centered ellipsoids for the counting regions, hoping to take advantage of set operations to generate unions of elliptical counting regions where necessary
to encompass an overall fragment.
While this scheme sounds more promising, the data presented in Figure \ref{fig:figure7} show that in practice, such set operations do not work out cleanly.
The trouble in this case is due to the fact that the optimal switching functions are much smoother than sharp step functions (which would have dire kinetic energy consequences),
and so clean set operations to create a union of neighboring counting regions are not achievable within the chosen 4-body form of the overall NCJF.
While the symmetry of ethene still allows for the eliminate ionic terms via an $\bm{F}$ that suppresses terms in which the left-hand and right-hand fragments'
not-quite-correctly-unioned counting regions give differing electron counts, the imperfections in the union create residual Jastrow curvature in between the
C and H atoms where the wave function magnitude is not small.
This residual curvature increases the kinetic energy of the neutral terms that survive the stenciling process, and it appears from the results in Figure \ref{fig:figure7}
that this effect is large enough that the variational principle instead chooses to eschew strong suppression of ionic terms.

The difficulties in the above two schemes highlight the properties that should be sought in future for general-purpose NCJF basis functions.
First, the function should be finite in spatial extent, so that when used for stenciling in one region they do not unduly affect the kinetic energy in distant
parts of the molecule.
Second, the functions must be capable of clean set operations so that they can combine when necessary to form a counting region around a group of atoms.
Finally, they must remain efficiently evaluable for a randomly chosen configuration of the electrons so as not to disrupt the algorithmic requirements
of VMC.
Although the presently tested planes and ellipsoids do not meet all of these requirements, the success of fragment-encompassing counting regions
in the challenging dissociation of ethene provides strong motivation to search for a formulation that does.

\section{Conclusions}
\label{sec:conclusions}

We have demonstrated that 4-body real space Jastrow factors are, with a suitable choice of Jastrow basis functions,
capable of performing strong wave function stenciling, in which a multiplicative Jastrow factor makes a large change to
the wave function by deleting unphysical configurations from a simple but overabundant fermionic reference.
In particular, these Jastrow factors are capable of eliminating ionic terms between well-separated molecular fragments,
which restores exact size consistency to the geminal power and greatly improves the situation for restricted Slater determinants,
bringing real space Jastrows in line with the size-consistency-restoring properties already enjoyed by Hilbert space Jastrows.
Unlike their Hilbert space brethren, the real space Jastrows presented here are compatible with diffusion Monte Carlo,
which creates exciting possibilities for generating qualitatively correct nodal surfaces in strongly correlated
regimes with a variational Monte Carlo approach that is both polynomial cost and size consistent.
Indeed, our preliminary results show that, when equipped with these stenciling-capable Jastrow factors, the variational
minimization of a single reference Jastrow-Slater trial function produces a qualitatively correct nodal surface
during the double bond dissociation of ethene, which in turn leads diffusion Monte Carlo to produce an accurate potential energy curve.
As every step in this process has a polynomially scaling cost, it will be very exciting in future to test the
efficacy of this combination in larger and more strongly correlated settings.

The key development allowing for effective stenciling was the introduction of a new form of 4-body Jastrow basis
function, in which a smoothed indicator function is used to check whether or not each electron is within a given
region of space.
These basis functions thereby allow the overall Jastrow factor to count and control how many electrons are in
a given region, which in turn allows for the suppression of unwanted ionic configurations.
Unlike previously explored basis function forms, these counting functions have no curvature except at the boundary
of their spatial region, allowing them to participate in strong stenciling so long as the boundaries are arranged so as
to hide their curvature in regions of small wave function magnitude.
In contrast, Gaussian-type basis functions have significant curvature at their centers, leading to
kinetic energy changes that prevent effective stenciling.
The most pressing priority in the future development of these number counting Jastrow factors is to formulate them
in a way that permits for black-box treatments of arbitrary molecules in which the variational principle
can decide automatically how to demarcate important regions in which to count and control electron number.
Although the planar and elliptical forms used in this study do not appear to support this black-box ideal, research
into promising alternatives is underway.

\section{Acknowledgements}
\label{sec:acknowledgements}

We acknowledge funding from the Office of Science, Office of Basic
Energy Sciences, the US Department of Energy, Contract No. DE-AC02-05CH11231.
Calculations were performed using the Berkeley Research Computing Savio cluster. 

\bibliographystyle{aip}

\end{document}